\documentclass[english,amsmath,amssymb,amsfonts,pra,twocolumn,showpacs,superscriptaddress,titlepage]{revtex4}

\usepackage[T1]{fontenc}
\usepackage[latin9]{inputenc}
\usepackage{graphicx}
\usepackage{bm}
\usepackage{ae,aecompl}
\usepackage{babel}
\begin{document}

\title{Degenerate Perturbation Theory Describing the Mixing of Orbital Angular
Momentum Modes in Fabry-P\'{e}rot Cavity Resonators}

\author{David H.~Foster}

\affiliation{Deep Photonics Corporation, Corvallis, Oregon 97333, USA}

\email{davidhfoster@gmail.com}

\author{Andrew K.~Cook}

\affiliation{Department of Physics, University of Oregon, Eugene, Oregon 97403,
USA}

\author{Jens U.~N\"{o}ckel}

\affiliation{Department of Physics, University of Oregon, Eugene, Oregon 97403,
USA}

\date{\today}

\begin{abstract}
We present an analytic perturbation theory which extends the paraxial approximation for a common cylindrically symmetric stable optical resonator and incorporates the differential, polarization-dependent reflectivity of a Bragg mirror. The degeneracy of Laguerre-Gauss modes with distinct orbital angular momentum (OAM) and polarization, but identical transverse order $N$, will become observably lifted at sufficiently small size and high finesse. The resulting paraxial eigenmodes possess two distinct OAM components, the fractional composition subtly depending on mirror structure. 
\end{abstract}


\maketitle
Polarization-dependent effects in three-dimensional optical systems
have in recent years received attention under the aspect of orbital
angular momentum (OAM) \cite{AllenProgOpt1999}. 
It has become important to understand the OAM interactions
with interfaces \cite{HostenSci2008, LoudonPRA2003}, waveguides \cite{DooghinSOcoupling} and resonators. 
One particular mechanism that warrants investigation is that of small corrections to the paraxial theory of resonators giving rise to what may be regarded as optical spin-orbit coupling \cite{FosterOL2004,BliokhOC2005}.

In this Rapid Communication, we extend electromagnetic resonator theory \cite{YuIEEETMTT1984} to provide a complete perturbation analysis and numerical
computations for an optical cavity which has an axis of rotational
symmetry $(\hat{\mathbf{z}})$, but nevertheless does not conserve
the component $\ell$ of OAM along that axis. 
This phenomenon itself
is remarkable because the model system, shown in Fig.\ 1(a), approaches
the \emph{paraxial} limit in which OAM conservation might be taken
for granted. 
Standard paraxial modes may be chosen to have well-defined $\ell$ because the polarization is transverse to $\hat{\mathbf{z}}$ and factors out of the wave problem, leaving a scalar Helmholtz equation which maps to
a quantum harmonic oscillator \cite{Nockel:07}.
Labeling the resulting transverse spectrum by $\ell\in\mathbb{Z}$
and a radial node number $p\in\mathbb{N}_{0}$, all modes with the
same transverse order $N=2p+|\ell|$ are degenerate in this set of approximations.
Therefore, going beyond this theory entails a degenerate perturbation
theory for which we construct a coupling Hamiltonian $V$ with the nominal Gaussian divergence angle, $\theta_{\text{D}}\equiv kw_{0}/2$, as its small parameter. 
Here $k$ is the wave number and $w_{0}$ is the waist radius.
Since $V$ is typically not diagonal in $\ell$, the solutions may be far from $\ell$-eigenstates. This holds even for arbitrarily small $\theta_{\text{D}}$.
Both polarization and $\ell$ mix in a paraxial spin-orbit coupling \cite{FosterOL2004,BliokhOC2005}.
Our work, in part, yields a new way of generating OAM and other non-uniformly polarized light.
Furthermore, in work involving cavity quantum electrodynamics, paraxial theory at the level developed here may be needed to distinguish spectral anti-crossings arising from passive cavity physics from those of strong photon-electron coupling. 

\begin{figure}
\includegraphics[width=1\columnwidth]{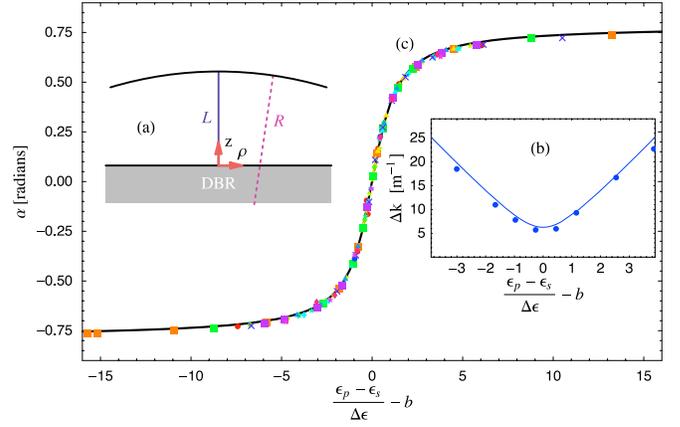}\caption{(a) Plano-concave model geometry. The top mirror radius of
curvature $R$ is $100\,\mu$m in the examples. 
(b) Wave number splitting $\Delta k \equiv k_{2}-k_{1} = \delta k_2 - \delta k_1$ from numerical
data at $L = 25\,\mu$m, $\lambda \approx 400\,$nm, indicating the avoided crossing between a mixable pair of modes.
(c) Numerical
results (symbols) for the mode mixing angle $\alpha$ closely fit the (solid)
curve $\alpha=\arctan[(\epsilon_{\text{p}}-\epsilon_{\text{s}})/\Delta\epsilon-b]/2$.
Different symbols (colors) represent data series at
$L=2.5$ to $25\,\mu$m (in steps of $2.5\,\mu$m) with $kR\approx780$ to $1631$.
$\Delta\epsilon$ and $b$ are fit here for each series.}
\end{figure}

The main result of our work is an analytical, quantitative expression for the degree of OAM mixing in a dome-shaped cavity with a Bragg mirror. 
The latter is crucial to achieving high finesse in optical microcavities. 
Our results also shed new light on the recently-observed
coupling between spatial and polarization degrees of freedom in broad-area
vertical-cavity surface emitting lasers \cite{babushkin:213901}:
polarization mixing was found to cause surprising spectral complexity even
with simple (square) boundaries, accompanied by intricate polarization
patterns in the far-field emission. 
The essential physics is provided by the interaction of differently polarized
plane-wave components with the planar, high-reflectivity distributed
Bragg reflector (DBR) on which both Ref.\ \cite{babushkin:213901}
and our system in Fig.\ 1 (a) are based.

Typical DBRs comprise dielectric multilayers and have a {}``form
birefringence'': the reflectivities for TE (s-polarized) and TM (p-polarized)
plane waves, $r_{\text{s(p)}}= |r_{\text{s(p)}}| \exp(i\phi_{\text{s(p)}})$, are unequal in phase
at nonzero angles of incidence, $\theta$. Previously \cite{FosterOL2004, FosterOC2004methods},
we had found numerically that the reflection phase difference $\phi_{\text{s}}(\theta_{\text{D}})-\phi_{\text{p}}(\theta_{\text{D}})\ne0$
is responsible for the polarization mixing. A consequence of great
practical importance is that by tuning cavity or mirror parameters,
the coupling of OAM imposed by Maxwell's equations can be
rigorously \emph{turned off} for selected modes, giving them well-defined
$\ell$. 
Aside from the consideration of OAM, our perturbation theory is fundamental
to paraxial theory itself: we perform a significant extension to and
completion of previous work by Yu and Luk \cite{YuIEEETMTT1984} which
derived the lowest order corrections to the paraxial modes of two-mirror cavities having perfect electrically conducting mirrors,
$r_{\text{s}}=r_{\text{p}}=-1$. Our approach, which allows a more general,
dielectric planar mirror, requires the construction of a $2 \times 2$ perturbation matrix $V$ and predicts very different results when $r_{\text{s}}$ and $r_{\text{p}}$ are different functions. 

The work of Ref.~\cite{YuIEEETMTT1984} and its precursors (e.g.~Ref.~\cite{EricksonIEEETMTT1975}) appears to be forgotten in the current literature.
This can be attributed to the difficulty of experimentally observing
the small spectral splittings caused by slightly non-paraxial perturbations. 
Agreement with a degenerate perturbation theory for $r_{\text{s(p)}}=-1$
was experimentally demonstrated using microwaves \cite{EricksonIEEETMTT1975}
where, due to comparable resonator size and wavelength, the mode spacing is large enough to resolve the lifting of the
$N+1$-fold degeneracies labeled by $N$. 
However, with recent progress
in miniaturization \cite{Cui:06}, comparable size parameters are becoming accessible
to optical cavities similar to Fig.\ 1(a). 

To write the splittings described in Ref.~\cite{YuIEEETMTT1984} in a
form we can use for the following discussion, let us first review
the unperturbed vector-field basis spanning the $N$-th transverse
multiplet. These are the \emph{Laguerre-Gauss} (LG) modes, written
in polar coordinates as a product $M_{N}^{\ell}(\rho,\phi,z)\hat{\bm{\sigma}}_{s}$
of a scalar part $M_{N}^{\ell}$, having orbital angular momentum $\ell$, and a circular polarization
vector $\hat{\bm{\sigma}}_{s}\equiv\left(\hat{\mathbf{x}}+si\hat{\mathbf{y}}\right)/\sqrt{2}$,
where $s=\pm1$ is the spin degree of freedom. 
Expressions for the LG modes and their Bessel wave decompositions are given in Ref.~\cite{FosterThesis}.
Perturbations that preserve rotational symmetry around $\hat{\mathbf{z}}$
will only couple those $M_{N}^{\ell}(\rho,\phi,z)\hat{\bm{\sigma}}_{s}$
for which the \emph{total angular momentum} $m=\ell+s$ around this
axis is the same \cite{FosterThesis}. 
We henceforth consider $N$
and $m$ to be fixed, nonnegative parameters; we will briefly discuss later the allowance of $m < 0$. The non-conservation
of OAM emerges here because $s$ can generally
\footnote{Specifically, when $N \geq 2$ and $0 < |m| <  N+1$, there are two mixable basis states and Eq.\ (\ref{eq: mixing}) applies.
} take two values, corresponding
to the basis states $M_{N}^{m-1}\hat{\bm{\sigma}}_{1}$ and
$M_{N}^{m+1}\hat{\bm{\sigma}}_{-1}$. 
No symmetry prevents these states from coupling, unless $m=0$ in which case time-reversal invariance applies. 

Nevertheless, no such OAM coupling is found in Ref.~\cite{YuIEEETMTT1984},
where perturbations merely split the wave numbers of the pair
 $\{M_{N}^{m\mp1}\hat{\bm{\sigma}}_{\pm1}\}$
by an amount
\begin{align}
\Delta k_{\text{LG}} \equiv k(M_{N}^{m+1}\hat{\bm{\sigma}}_{-1}) - k(M_{N}^{m-1}\hat{\bm{\sigma}}_{1}) = \frac{m}{4 k L R}.\label{eq: YuLuk splitting}
\end{align}
The calculation assumes a dome of vertical length $L$, top mirror
radius of curvature $R$, and $r_{\text{s(p)}} = -1$. We will
now show that a more realistic model for $r_{\text{s(p)}}$ leads to
actual resonator modes $\bm{\Psi}_{N,m,1}$ and $\bm{\Psi}_{N,m,2}$
having a given $m$ but forming a rotation of the LG basis pair by
a mixing angle $\alpha\in(-\pi/4,\pi/4)$:
\begin{align}
\begin{pmatrix}\bm{\Psi}_{N,m,1}(\alpha)\\
\bm{\Psi}_{N,m,2}(\alpha)\end{pmatrix}=\begin{pmatrix}\cos\alpha & -\sin\alpha\\
\sin\alpha & \cos\alpha\end{pmatrix}\begin{pmatrix}M_{N}^{m-1}\hat{\bm{\sigma}}_{1}\\
M_{N}^{m+1}\hat{\bm{\sigma}}_{-1}\end{pmatrix}.\label{eq: mixing}
\end{align}
The states at $\alpha = \pm\pi/4$ are \emph{hybrid modes}, one of which is predominantly (though not completely) composed of TM plane waves, the other being predominantly TE \cite{FosterOL2007, FosterThesis}. 
This approximate polarization separation of the hybrid modes yields an intuitive picture of mode mixing; the separation acts as a lever arm by which the Bragg mirror, having $\phi_{\text{s}} \neq \phi_{\text{p}}$, rotates the eigenmode basis away from the LG modes and toward the hybrid modes.

The OAM-non-conservation described above is exhibited by the paraxial
modes of two-mirror axisymmetric cavity resonators which 1) are of
sufficiently small size with respect to wavelength, 2) have sufficiently
narrow resonance widths (low loss), and 3) have at least one mirror
for which $\phi_{\text{s}} \neq \phi_{\text{p}}$ (commercial dielectric
mirrors meet this requirement). 
The first two requirements are essential
in splitting the degeneracies of high order Gaussian modes, and
Eq.\ (\ref{eq: YuLuk splitting}) allows us to estimate whether this
is possible. To perform the degenerate perturbation theory in the
presence of property 3), we first develop the perturbation
Hamiltonian, $V$, in the two-mode basis $\{M_{N}^{m\mp1}\hat{\bm{\sigma}}_{\pm1}\}$.
$V$ must be symmetric $(V_{21}=V_{12})$, and for our purposes may
be taken to be traceless $(V_{22}=-V_{11})$.
$V$ then has eigenvectors $\vec{v}_1 = \bigl( \begin{smallmatrix} \cos \alpha \\ - \sin \alpha \end{smallmatrix} \bigr), \vec{v}_2 = \bigl( \begin{smallmatrix} \sin \alpha \\ \cos \alpha \end{smallmatrix} \bigr)$ with
\begin{align}
\alpha = (1/2) \arctan(V_{12} / V_{22}),
\end{align} 
and eigenvalues $\delta k_{1(2)}=\mp(V_{12}^{2}+V_{22}^{2})^{1/2}$.
As system parameters are varied, the elements of $V$ change and an anti-crossing of the hybrid modes emerges, cf.~Fig.\ 1(b). 

All deviations from the paraxial limit must be considered to lowest
order in $\theta_{\text{D}}^2$, which for our cavity is $\theta_{\text{D}}^{2} = 2/ [k \sqrt{L(R-L)} ]$.
The physical derivation of $V$ is facilitated by noting that in
the anti-crossing scenario, $\alpha=0$ is equivalent to $V_{12}=0$
and hence corresponds to the assumptions underlying the known result
Eq.\ (\ref{eq: YuLuk splitting}). Thus, Eq.\ (\ref{eq: YuLuk splitting})
should be reproduced by our model at $\alpha=0$. In wave number units,
we therefore set
\begin{align}
V_{22}=-V_{11}=\frac{1}{2}\Delta k_{\text{LG}}=\frac{m}{16}\sqrt{\frac{L}{R}\left(1-\frac{L}{R}\right)}\,\frac{\theta_{\text{D}}^{2}}{L}. \label{eq: V diagonal}
\end{align}
Although Eq.\ (\ref{eq: YuLuk splitting})
was derived for ideal-metal cavities, our more general DBR boundary
conditions do not affect Eq.\ (\ref{eq: V diagonal}), because they
shift all LG modes with the same $N$ \emph{equally}. To explain this,
consider the \emph{penetration depths} $\delta L_{\text{s(p)}}$ of each of $M_{N}^{m\mp1}\hat{\bm{\sigma}}_{\pm1}$
into the mirror layers. For a plane wave of s or p polarization with
incident angle $\theta$ at a DBR, $\delta L_{\text{s(p)}} = \phi_{\text{s(p)}}(k,\theta)/(2k)$.
In order to capture the relevant material properties of the DBR, we neglect transmission and expand its reflection phase in the plane wave angle of incidence,
$\theta$, as
$\phi_{\text{s(p)}}(k,\theta)\approx\phi_{0}(k)+\epsilon_{\text{s(p)}}(k)\theta^{2}$.

Any vectorial mode $\bm{\Psi}$ can be decomposed into azimuthally symmetrized plane waves (Bessel waves) by defining a {}``tilde'' operator
such that $\tilde{\Psi}^{\text{s(p)}}(\theta)$ essentially denotes the amplitudes of the TE(TM) plane waves of polar angle $\theta$.
The paraxial $\tilde{\Psi}^{\text{s(p)}}(\theta)$ is nonzero
only near $\theta\approx0$, with the \emph{averaged} reflection
phase of $\bm{\Psi}$ being
\begin{equation}
\left\langle \phi\right\rangle _{\bm{\Psi}}\equiv\int\left[ \bigl| \tilde{\Psi}^{\text{s}}(\theta)\bigr|^{2}\,\phi_{\text{s}}(\theta)+\bigl|\tilde{\Psi}^{\text{p}}(\theta)\bigr|^{2}\,\phi_{\text{p}}(\theta)\right] \theta\, d\theta,\label{eq:GeneralPWaverage}
\end{equation}
with the normalization $\int(\vert\tilde{\Psi}^{\text{s}}\vert^{2}+\vert\tilde{\Psi}^{\text{p}}\vert^{2})\theta\, d\theta=1$.
Specializing to the LG modes with the abbreviation $\bm{\Lambda}_{\pm1}\equiv M_{N}^{m\mp1}\hat{\bm{\sigma}}_{\pm1}$,
circular polarization leads to $\vert\tilde{\Lambda}_{\pm1}^{\text{s}}(\theta)\vert^{2} = \vert\tilde{\Lambda}_{\pm1}^{\text{p}}(\theta)\vert^{2}$.
The harmonic-oscillator nature of the transverse field \cite{Nockel:07}
entails that Eq.\ (\ref{eq:GeneralPWaverage}), 
with the above expansion for $\phi_{\text{s(p)}}$, 
depends only on the mode order $N$
but not on $\ell$. This carries over to the
average penetration depth $\left\langle \delta L\right\rangle _{\bm{\Lambda}_{\pm1}}=\left\langle \phi\right\rangle _{\bm{\Lambda}_{\pm1}}/(2k)$,
and hence the effective cavity length $L+\left\langle \delta L\right\rangle _{\bm{\Lambda}_{\pm1}}$
is identical for both modes $\bm{\Lambda}_{\pm1}$; this then implies
equal spectral shifts, as claimed above.

To obtain the off-diagonal element $V_{12}$, we apply the same penetration-depth
argument to the special case $\alpha=\pi/4$ where Eq.\ (\ref{eq: mixing})
yields the hybrid modes. Their plane-wave amplitudes, $\tilde{\Psi}_{N,m,1}^{\text{s(p)}}(\theta)$
and $\tilde{\Psi}_{N,m,2}^{\text{s(p)}}(\theta)$, can be written
purely in terms of the LG amplitudes $\tilde{\Lambda}_{\pm1}^{\text{p}}(\theta)$
using Eq.\ (\ref{eq: mixing}), and their splitting, $\Delta k_{\text{hybrid}}\equiv k_{N,m,2}-k_{N,m,1}$, is given by 
\begin{eqnarray}
\Delta k_{\text{hybrid}} & = & -(k / L) \left(\delta L_{N,m,2}-\delta L_{N,m,1}\right)\label{eq:SplittingQTEM}\\
 & = & \frac{\epsilon_{\text{p}}-\epsilon_{\text{s}}}{L}\int\theta^{3}\tilde{\Lambda}_{+1}^{\text{p}}(\theta)\,\tilde{\Lambda}_{-1}^{\text{p}}(\theta)\, d\theta, \label{eq:SplittingQTEM0}
\end{eqnarray}
where $k$ is the unperturbed wave number. Therefore, we reach $\left|\Delta k_{\text{\text{hybrid}}}\right|\gg\left|\Delta k_{\text{LG}}\right|$
if the form birefringence quantity, $\left|\epsilon_{\text{p}}-\epsilon_{\text{s}}\right|$,
is made large. On the other hand, $\alpha \rightarrow \pi/4$ implies $\left|V_{12}\right|\gg\left|V_{22}\right|$,
so that the eigenvalues of $V$ in this limit are $\delta k_{1(2)}\approx\mp V_{12}$.
Setting $2 \delta k_{2}$ equal to Eq.\ (\ref{eq:SplittingQTEM0}) and
performing the $\theta$ integral, one obtains
\begin{align}
V_{12}=V_{21} & =\frac{\epsilon_{\text{p}}(k)-\epsilon_{\text{s}}(k)}{8}\sqrt{(N+1)^{2}-m^{2}}\,\frac{\theta_{\text{D}}^{2}}{L}.\label{eq: V off-diagonal}
\end{align}

For some simple dielectric mirrors, $\epsilon_{\text{p}}-\epsilon_{\text{s}}$
may be swept across zero by varying the cavity length across the nominal
$L$ at which the mode pair of interest has unperturbed $k$ equal
to the design (center) wave number of the mirror, $k_{\text{d}}$.
The mixing angle $\alpha$ can then be written as
\begin{align}
\alpha & = (1/2) \arctan \bigl\{ [ \epsilon_{\text{p}}(k)-\epsilon_{\text{s}}(k) ] / \Delta\epsilon \bigr\}, \label{eq: mixing angle}
\end{align}
 where the width of the crossover interval is given by
\begin{align}
\Delta\epsilon(L/R,N,m) & =\frac{m}{2} \sqrt{\frac{(L/R)(1-L/R)}{(N+1)^{2}-m^{2}}}. \label{eq: Delta epsilon}
\end{align}

 The formulas above complete the lowest order degenerate perturbation
theory for paraxial mode mixing. 
Interestingly, $\Delta\epsilon$ is \emph{independent of wavelength}: for fixed cavity geometry and mode labels $N$, $m$, modes of different
longitudinal node number (along $\hat{\mathbf{z}}$) will have different
$k$ but identical anti-crossing behavior when $\alpha$ is plotted
versus $\epsilon_{\text{p}}-\epsilon_{\text{s}}$. This universal
functional form provides a robust way of tailoring any desired mixing
angle $\alpha$. Most importantly under the aspect of OAM non-conservation,
we can tune Eq.\ (\ref{eq: mixing}) to $\alpha=0$.

Once the cavity linewidth for the relevant paraxial modes becomes
less than the mode separation, $\Delta k \equiv 2 \delta k_{2} \geq m / (4 k L R)$, the two modes $\bm{\Psi}_{N,m,j}$
given by Eqs.~(\ref{eq: mixing}, \ref{eq: mixing angle}, \ref{eq: Delta epsilon})
are resolved at slightly different $k$ (or $L$). Any excitation
of the cavity would generally have non-zero overlap with these modes,
and thus complicated mode patterns \cite{FosterOL2004, FosterThesis} can be generated by simple
excitation. 
The magnitude of $\alpha$ is zeroth order in $\theta_{\text{D}}^{2}$,
and excursions near the asymptotic values can be seen in Fig.~1(c). 
The magnitude of the relative frequency splitting, however, is $O(\theta_{\text{D}}^{4})$ as $\theta_{\text{D}} \rightarrow 0$.
We note that observed modes will not be restricted to $m \geq 0$. 
Axial symmetry creates an exact two-fold degeneracy of the vectorial LG basis modes under the transformation in which $m$, $\ell$, and $s$ switch sign \cite{FosterOC2004methods}. 
The presence of the exact degeneracy \footnote{
This results in a degenerate (superimposable) SU(2) sector which externally multiplies the non-degenerate (mixable) SU(2) sector we have considered in Eq.~(\ref{eq: mixing}).
The location, or ``generalized polarization'', of each observed mode within the exactly degenerate sector is excitation dependent, while the cavity fixes the location ($\alpha$) within the mixable sector.
} does not ``wash out'' the generation of complicated mode patterns and is more fully discussed in Refs.\ \cite{FosterOC2004methods, FosterThesis, NockelPRA2009}.

\begin{figure}
\includegraphics[width=1\columnwidth]{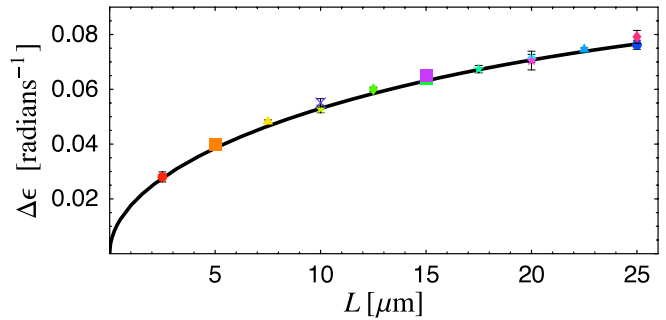}
\caption{Mode-coupling width $\Delta\epsilon$ from fit (symbols) compared to perturbative prediction (solid line),
plotted as a function of the cavity length $L$ at fixed $R=100\,\mu$m. 
Error bars were obtained from the fits of Fig.~1(c).
Because the analytic equation (\ref{eq: Delta epsilon}) is wavelength-independent,
all data for $\lambda\approx400\,$nm and $\lambda\approx800\,$nm
fall onto the same curve.}
\end{figure}

We have numerically calculated modes for $R=100\,\mu$m and
$L=2.5$ to $25\,\mu$m with a DBR comprising 36 pairs of quarter-wave
dielectric layers A and B with refractive indices $n_{\text{A}}=3.52$ and
$n_{\text{B}}=3.00$, with layer B at the top surface. 
Typical values of $\epsilon_{\text{s(p)}}$
were around $-3$, with $\text{d}(\epsilon_{\text{p}}-\epsilon_{\text{s}})/\text{d}(k-k_{\text{d}})\approx-2.8\, \mu$m. 
Data were analyzed for modes close to $\lambda_{\text{d}} \equiv 2 \pi / k_{\text{d}} \approx400$ and $800$ nm. 
We considered mode pairs with $N=2$ and $m=1$, the lowest values for which OAM mixing can occur.

The numerical data for the mixing angle are fit extremely
well by Eq.\ (\ref{eq: mixing angle}) if we allow for an offset $b$
in the argument of the arctan, as done in Fig.\ 1(c). The comparison
between numerical fit and Eq.~(\ref{eq: Delta epsilon}) is shown
in Fig.~2. 
The offset $b$ has median $-0.32$ for our data and empirically behaves as $C/[k L(1-L/R)]$,
where $C$ is a slowly-varying function of $k$ and $k_{\text{d}}$. 
Taking the limit $\theta_{\text{D}}\rightarrow 0$ such that $L/R$ is bounded away from 0 and 1 implies that $|b| = O(\theta_{\text{D}}^{2})$. 
This next-highest-order correction to our perturbation theory will be discussed in a subsequent publication \cite{NockelPRA2009}.

To spectrally resolve the transverse mode splitting along the entire mixing curve, the cavity must obey $2 k R [1- (\text{R}_1 \text{R}_2)^{1/2} ] < m$, where $\text{R}_{1(2)}$ are the power reflectivities of the two mirrors. 
Microwave experiments may be the most direct approach.
Alternatively, paraxial spin-orbit coupling may be realized in microcavity resonators for quantum-information applications, cf.~Ref.~\cite{Cui:06}, where small size and high finesse are required.
Such cavities could generate OAM or hybrid beams at light levels from single-photon-on-demand to that of a macroscopic laser.
In particular, one could utilize the fine structure and varied spatial patterns of the split modes. 
This has particular potential for quantum information applications: the order $N$ family provides $N+1$ nearly degenerate energy levels corresponding to modes
having different vectorial spatial patterns of the electric field. 
This spectral and spatial structure combined with quantum dots at the planar mirror may possess quantum logic capability. 

In conclusion, both the mixing angle and the frequency splitting for
OAM-mixed paraxial resonator modes in an axisymmetric cavity have been analytically derived here in a degenerate
perturbation theory which includes the form birefringence of a practical mirror.
The phenomenon is similar in principle to the polarization coupling observed in Refs.~\cite{babushkin:213901, FrattaPRA2001}: differences in the penetration depths for TE and TM plane waves modify the vectorial resonator modes.
For our case however, eigenmode coupling persists from the non-paraxial regime to the deeply paraxial regime. 
By examining the latter, we have shown here that the coupling can in fact be turned on and off via the material parameters $\epsilon_{\text{s(p)}}$.
The underlying inadequacy of a scalar paraxial treatment is washed out in macroscopic cavities, but must be regarded as a fundamental limitation in high-finesse microcavities, where the correct starting point for any paraxial formulation must be the mode basis of Eq.~(\ref{eq: mixing}), which is heterogeneous in orbital angular momentum.

This work is supported in part by National Science Foundation Grant
No.~ECS-0239332.


\end{document}